\documentclass[12pt]{JHEP3}

\usepackage{epsfig}

\def\spose#1{\hbox to 0pt{#1\hss}}
\def\ltapprox{\mathrel{\spose{\lower 3pt\hbox{$\mathchar"218$}}
 \raise 2.0pt\hbox{$\mathchar"13C$}}}
\def\gtapprox{\mathrel{\spose{\lower 3pt\hbox{$\mathchar"218$}}
 \raise 2.0pt\hbox{$\mathchar"13E$}}}

\title{
$\theta$ dependence of the spectrum of SU($N$) gauge theories
}

\author{Luigi Del Debbio \\
        SUPA, School of Physics, University of Edinburgh, 
        Edinburgh EH9 3JZ, UK\\
        E-mail: \email{luigi.del.debbio@ed.ac.uk} 
} 

\author{Gian Mario Manca \\
        Dipartimento di Fisica dell'Universit\`a di Parma and I.N.F.N., 
        I-43100 Parma, Italy \\ 
        E-mail: \email{manca@fis.unipr.it} 
} 

\author{Haralambos Panagopoulos \\ 
        Department of Physics, University of Cyprus,
        Lefkosia, CY-1678, Cyprus\\ 
        E-mail: \email{haris@ucy.ac.cy} 
}

\author{Apostolos Skouroupathis \\ 
        Department of Physics, University of Cyprus,
        Lefkosia, CY-1678, Cyprus\\ 
        E-mail: \email{php4as01@ucy.ac.cy} 
}

\author{Ettore Vicari \\ 
        Dipartimento di Fisica dell'Universit\`a 
        di Pisa and I.N.F.N., I-56127 Pisa, Italy \\ 
        E-mail: \email{vicari@df.unipi.it} 
}

\abstract{ We study the $\theta$ dependence of the spectrum of
  four-dimensional SU($N$) gauge theories, where $\theta$ is the coefficient
  of the topological term in the Lagrangian, for $N\ge 3$ and in the large-$N$
  limit.  We compute the $O(\theta^2)$ terms of the expansions around
  $\theta=0$ of the string tension and the lowest glueball mass, respectively
  $\sigma(\theta) = \sigma \left( 1 + s_2 \theta^2 + ... \right)$ and
  $M(\theta) = M \left( 1 + g_2 \theta^2 + ... \right)$, where $\sigma$ and
  $M$ are the values at $\theta=0$.  For this purpose we use numerical
  simulations of the Wilson lattice formulation of SU($N$) gauge theories for
  $N=3,4,6$.  The $O(\theta^2)$ coefficients turn out to be very small for all
  $N\ge 3$.  For example, $s_2=-0.08(1)$ and $g_2=-0.06(2)$ for $N=3$.  Their
  absolute values decrease with increasing $N$.  Our results are suggestive of
  a scenario in which the $\theta$ dependence in the string and glueball
  spectrum vanishes in the large-$N$ limit, at least for sufficiently small
  values of $|\theta|$. They support the general
  large-$N$ scaling arguments that indicate $\bar{\theta}\equiv \theta/N$ as
  the relevant Lagrangian parameter in the large-$N$ expansion.  }


\begin{document}

\section{Introduction}
\label{intro}

Four-dimensional SU($N$) gauge theories have a nontrivial dependence
on the angle $\theta$ that appears in the Euclidean Lagrangian as
\begin{equation}
{\cal L}_\theta  = {1\over 4} F_{\mu\nu}^a(x)F_{\mu\nu}^a(x)
- i \theta {g^2\over 64\pi^2} \epsilon_{\mu\nu\rho\sigma}
F_{\mu\nu}^a(x) F_{\rho\sigma}^a(x)
\label{lagrangian}
\end{equation}
($q(x)=\frac{g^2}{64\pi^2} \epsilon_{\mu\nu\rho\sigma}
F_{\mu\nu}^a(x) F_{\rho\sigma}^a(x)$ is the topological
charge density).
Indeed, the most plausible explanation of how the solution of the
so-called U(1)$_A$ problem can be compatible with the $1/N$ expansion
(performed keeping $g^2N$ fixed \cite{Hooft-74}) requires a nontrivial
$\theta$ dependence of the ground-state energy density $F(\theta)$, 
\begin{equation}
F(\theta) = -{1\over V} \ln \int [dA] \exp \left(  - \int d^d x {\cal L}_\theta
\right),
\label{GSE}
\end{equation}
in the $d$-dimensional pure gauge theory to leading order in
$1/N$~\cite{Witten-79,Veneziano-79}. 
The large-$N$ ground-state energy is expected to behave as
\cite{Witten-80,Witten-98,Gabadadze-99}
\begin{equation}
\Delta F(\theta)\equiv F(\theta) - F(0) = {\cal A} \, \theta^2  +
O\left( 1/N^2\right) 
\label{conj}
\end{equation} 
for sufficiently small values of $\theta$, i.e. $\theta<\pi$.  This
has been supported by Monte Carlo simulations of the 
lattice formulation of SU($N$) gauge theories \cite{DPV-02}. 
Indeed, the numerical results for $N=3,4,6$ are consistent with
a scaling behavior around $\theta=0$  given by
\begin{eqnarray}
&&f(\theta) \equiv \sigma^{-2} \Delta F(\theta) 
= {1\over 2} C \theta^2 ( 1 + b_2 \theta^2 + ...), 
\label{ftheta} \\
&& C=C_\infty + c_2/N^2 + ... , \qquad b_2=b_{2,2}/N^2+...
\nonumber
\end{eqnarray}
where $\sigma$ is the string tension at $\theta=0$.
$C$ is the ratio
$\chi/\sigma^2$ where $\chi = \int d^4 x \langle q(x)q(0) \rangle$ 
is the topological susceptibility at $\theta=0$.
Its large-$N$ limit $C_\infty$ is \cite{DPV-02,LT-01t}
$C_{\infty}\approx 0.022$. Moreover, estimates of 
$c_2$ and $b_{2,2}$ are \cite{DPV-02} $c_2\approx
0.06$, and $b_{2,2}\approx -0.2$
($b_2\approx -0.02$ for SU(3) \cite{DPV-02,Delia-03}).
Note that Eq.~(\ref{ftheta}) can be recast in the form
\begin{eqnarray} 
&&f(\theta) = N^2 \bar{f}(\bar{\theta}\equiv \theta/N), 
\label{fthetabar} \\
&& \bar{f}(\bar{\theta}) = 
{1\over 2} C \bar{\theta}^2 ( 1 + \bar{b}_2 \bar{\theta}^2 + ...), 
\nonumber
\end{eqnarray}
where $\bar{b}_2=b_{2,2}+ O(1/N^2)=O(1)$.  This is consistent with general
large-$N$ scaling arguments applied to the Lagrangian (\ref{lagrangian}),
which indicate $\bar{\theta}\equiv \theta/N$ as the relevant Lagrangian
parameter in the large-$N$ limit of the ground-state
energy \cite{Witten-80}.

Another interesting issue concerns the $\theta$ dependence of the
spectrum of the theory. This is particularly interesting in the
large-$N$ limit where the issue may also be addressed by other approaches,
such as AdS/CFT correspondence applied to nonsupersymmetric and non
conformal theories, see e.g. Ref.~\cite{AGMOO-00}. The analysis  of 
the $\theta$ dependence of the glueball spectrum using AdS/CFT
suggests that the only effect of the $\theta$ term in the leading
large-$N$ limit is that the lowest spin-zero glueball state becomes a
mixed state of $0^{++}$ and $0^{-+}$ glueballs, as a consequence of
the fact that the $\theta$ term breaks parity, but its mass does not
change \cite{GI-04}.

In this paper we present an exploratory numerical study of the $\theta$
dependence in the spectrum of SU($N$) gauge theories. For this purpose we use
numerical simulations of the Wilson lattice formulation. Numerical Monte Carlo
studies of the $\theta$ dependence are made very difficult by the complex
nature of the $\theta$ term.  In fact the lattice action corresponding to the
Lagrangian (\ref{lagrangian}) cannot be directly simulated for $\theta\ne 0$.
Here we restrict ourselves to the region of relatively small
$\theta$ values, where one may expand the observable values around $\theta=0$.
We consider the string tension and the lowest glueball mass.  
We write
\begin{equation}
\sigma(\theta) = \sigma \left( 1 + s_2 \theta^2 + ... \right),
\label{sigmaex}
\end{equation}
where $\sigma$ is the string tension at $\theta=0$.
When $N\ge 4$ analogous expressions can be written for the other
independent $k$-strings associated with group representations
of higher $n$-ality.
Moreover, for the lowest glueball state we write
\begin{equation}
M(\theta) = M\left( 1 + g_2 \theta^2 + ... \right)
\label{gmex}
\end{equation} 
where $M$ is the $0^{++}$ glueball mass at $\theta=0$.  Then the coefficients
of these expansions can be computed from appropriate correlators at
$\theta=0$.  The $O(\theta^2)$ coefficients $s_2$ and $g_2$ are dimensionless
quantities, which should approach a constant in the continuum limit, with
$O(a^2)$ scaling corrections.  The idea is analogous to the one exploited in
Ref.~\cite{DPV-02} to study the $\theta$-dependence of the ground-state
energy.

We shall present results for four-dimensional SU($N$) gauge theories with
$N=3,4,6$.  The estimates of the $O(\theta^2)$ coefficients turn out to be
very small for all $N\ge 3$.  For example $s_2=-0.08(1)$ and $g_2=-0.06(2)$
for $N=3$.  Moreover, their absolute values decrease with increasing $N$. We
also observe that the $O(\theta^2)$ terms are substantially smaller in
dimensionless ratios such as $M/\sqrt{\sigma}$ and, for $N>3$, the ratios of
independent $k$ strings, $R_k=\sigma_k/\sigma$.  Our results are suggestive of
a large-$N$ scenario in which the $\theta$ dependence in the string and
glueball spectrum vanishes around $\theta=0$. They are consistent with the
general large-$N$ scaling arguments indicating $\bar{\theta}\equiv \theta/N$
as the relevant parameter in the large-$N$ limit.  We also show that a similar
scenario emerges in the two-dimensional CP$^{N-1}$ models by an analysis of
their $1/N$ expansion.

The paper is organized as follows.  In Sec.~\ref{sec2} we outline the
numerical method to estimate the $O(\theta^2)$ terms of the expansion
in powers of $\theta$ around $\theta=0$. The results of our
exploratory numerical study are presented in Sec.~\ref{res}. Finally,
in Sec.~\ref{cpn} we discuss the $\theta$ dependence of
two-dimensional CP$^{N-1}$ models within their $1/N$ expansion around
their large-$N$ saddle-point solution.

\section{Numerical method}
\label{sec2}

\subsection{Monte Carlo simulations}
\label{mc}

We consider the Wilson formulation of lattice gauge theories:
\begin{equation}
S = - N\beta \sum_{x,\mu>\nu} {\rm Tr} \left[
U_\mu(x) U_\nu(x+\mu) U_\mu^\dagger(x+\nu) U_\nu^\dagger(x) 
+ {\rm h.c.}\right],
\label{wilsonac}
\end{equation}
where $U_\mu(x)\in$ SU($N$) are link variables.
In our simulations we employed the Cabibbo-Marinari algorithm
\cite{CM-82} to upgrade SU($N$) matrices by updating their SU(2)
subgroups (we selected $N(N-1)/2$ subgroups and each matrix upgrading
consists of $N(N-1)/2$ SU(2) updatings).  This was done by alternating
microcanonical over-relaxation and heat-bath steps, typically in a 4:1
ratio.

Computing quantities related to topology using lattice simulation techniques
is not a simple task.  In a lattice theory the fields are defined on a
discretized set, therefore the topological properties are strictly trivial.
One relies on the fact that the physical topological properties are recovered
in the continuum limit.  Various techniques have been proposed and employed to
associate a topological charge $Q$ to a lattice configuration, see, e.g, 
Refs.~\cite{CDPV-90,Teper-00} for techniques based on bosonic operators, and
Refs.~\cite{DP-04,GLWW-03} for techniques based on fermionic estimators.
The most robust definition of topology on the lattice is
the one obtained using the index
of the overlap Dirac operator. However, due to the computational cost 
of fermionic methods and the need for very large statistics to measure 
correlations of Polyakov and plaquette operators with topological quantities,
we decided to use the simpler cooling method, implemented as in Ref.~\cite{DPV-02}. 
Direct comparison with a fermionic estimator is known to show a good agreement
in the case of SU(3)~\cite{DP-04,DPV-02,DGP-05}, supporting the idea that the cooling
method is fairly stable in this case.  Moreover,
the agreement among different methods is expected to improve with increasing $N$
\cite{RRV-97,CTW}.

A severe form of critical slowing down affects the measurement of $Q$, posing
a serious limitation for numerical studies of the topological properties in
the continuum limit, especially at large values of $N$.  The autocorrelation
time $\tau_Q$ of the topological modes rapidly increases with the length
scale, much faster than the standard square law of random walks
\cite{DPV-02,DMV-04}.  The available estimates of $\tau_Q$ appear to increase
as an exponential of the length scale, or with large power laws.  A
qualitative explanation of this severe form of critical slowing down may be
that topological modes give rise to sizeable free-energy barriers separating
different regions of the configuration space.  As a consequence, the evolution
in configuration space may present a long-time relaxation due to transitions
between different topological charge sectors.  This dramatic effect has not
been observed in plaquette-plaquette or Polyakov line correlations, suggesting
an approximate decoupling between topological modes and nontopological ones,
such as those determining the confining properties and the glueball spectrum.
But, as we shall see, such a decoupling is not complete. Therefore the strong
critical slowing down that is clearly observed in the topological sector will
eventually affect also the measurements of nontopological quantities, such as
those related to the string and glueball spectrum.

\subsection{The $O(\theta^2)$ coefficients of the $\theta$ expansion}
\label{thex}

In this subsection we describe the method to determine the
$O(\theta^2)$ coefficients of the $\theta$ expansions such as
Eqs.~(\ref{sigmaex}) and (\ref{gmex}).
Let us first discuss the case of the fundamental string tension.  
The string tension can be determined from the torelon mass, i.e. the
mass describing the large-time exponential 
decay of the wall-wall correlations $G_P$ of Polyakov lines \cite{DSST-85}.  
In the presence of a $\theta$ term, 
\begin{equation}
G_P(t,\theta)=\langle A_P(t) \rangle_\theta = 
{ \int [dU] A_P(t) e^{-\int d^4 x {\cal L}_\theta}
\over \int [dU] e^{-\int d^4 x {\cal L}_\theta }}
\end{equation}
where
\begin{equation}
A_P(t) = \sum_{x_1,x_2}  {\rm Tr}\,P(0;0) \; {\rm Tr}\,P(x_1,x_2;t),  
\label{apdef}
\end{equation}
and $P(x_1,x_2;t)$ is the Polyakov line along the $x_3$ direction of
size $L$.
The time separation $t$ is an integer multiple of the lattice spacing
$a$: $t = n_t\,a$.
The correlation $G_P$ can be expanded in powers of $\theta$. 
Here, we are considering the case of the fundamental
string tension, but the discussion can be easily extended
to any other group representation, by replacing 
the trace with the corresponding character in Eq.~(\ref{apdef}).
Taking into account the parity symmetry at $\theta=0$, we obtain
\begin{equation}
G_P(t,\theta) = G_P^{(0)}(t) + \frac{1}{2}\theta^2 G_P^{(2)}(t) + O(\theta^4),
\end{equation}
where 
\begin{eqnarray}
&& G_P^{(0)}(t) = \langle A_P(t) \rangle_{\theta=0},\\
&& G_P^{(2)}(t) = - \langle A_P(t) Q^2 \rangle_{\theta=0} + 
\langle A_P(t) \rangle_{\theta=0} \langle Q^2 \rangle_{\theta=0}
\label{g2}
\end{eqnarray}
and $Q$ is the topological charge.

The correlation function $G_P$ is expected to have
a large-$t$ exponential behavior
\begin{equation}
G_P(t,\theta) \approx B(\theta) e^{-E(\theta) t},
\label{gptlarge}
\end{equation}
where  $E(\theta)$ is the $\theta$-dependent energy of the lowest state
(torelon mass), and $B(\theta)$ is the overlap of the 
source with the lowest-energy state.
If the lattice size $L$ is sufficiently large, the lowest-energy states
describing the  long-distance behavior of
Polyakov correlators should be those of a string-like spectrum.
Then, the string tension is extracted using the relation 
\begin{equation}
E(\theta) = \sigma(\theta) L - {\pi\over 3L} 
\label{lute}
\end{equation}
Here we are assuming that the $O(1/L)$ (L\"uscher) correction is 
independent of $\theta$.
Actually we are also assuming the so called free string spectrum 
\begin{equation}
W_n = \sigma L \left(1 - {\pi \over 3 l_\sigma^2} + n{4\pi\over
    l_\sigma^2}\right),
\qquad
l_\sigma\equiv  \sqrt{\sigma}L 
\label{wnfree}
\end{equation}
($n$: excitation level), obtained neglecting the self-interaction
terms in the string effective action, see e.g. Ref.~\cite{LW-04}. As
shown in Ref.~\cite{LW-04}, only the 
$O(1/L)$ correction should be 
universal, while subleading corrections are generally expected. They depend on
the unknown coefficients of the higher order terms of the effective QCD string
action.  For example, besides the free string spectrum, one may also consider
the Nambu-Goto string spectrum~\cite{LW-04,Arvis}
\begin{equation}
W_n = \sigma L \left( 1  - {2\pi \over 3 l_\sigma^2} + 
n{8\pi\over l_\sigma^2}\right)^{1/2}
\label{wnNG}
\end{equation}
In particular, if one assumes the Nambu-Goto spectrum, instead of
Eq.~(\ref{lute}), one should use the Nambu-Goto lowest-energy state 
to determine the string tension, i.e.
\begin{equation}
W_0 = \sigma L \left( 1  - {2\pi \over 3 l_\sigma^2} \right)^{1/2}=
\sigma L\left[ 1 - {\pi \over 3 l_\sigma^2} -  
{\pi^2 \over 18 l_\sigma^4 } + O\left( {1/l_\sigma^6 }\right)\right] 
\label{wnNG0}
\end{equation}
Therefore the Nambu-Goto string spectrum leads
to a different estimate of the string tension at finite $l_\sigma$:
\begin{equation}
\sigma_{\rm fs}=\sigma_{\rm NG} \left[ 1 - 
{\pi^2 \over 18 l_\sigma^4 } + O(l_\sigma^{-6})\right] 
\end{equation}
where $\sigma_{\rm fs}$ and $\sigma_{\rm NG}$ are the string tensions
extracted from the torelon mass assuming respectively the free and
Nambu-Goto spectrum.  

Lattice sizes such that $l_\sigma\gtrsim 3$ should be sufficiently large to
have an effective string picture parametrized by a constant string tension
$\sigma$ \cite{LT-01}.  Then the different estimates of $\sigma$ obtained by
using the free and Nambu-Goto spectra might provide an estimate of
systematic error on the determination of $\sigma$ from the lowest torelon mass
due to our partial knowledge of the effective QCD string action.  For
$l_\sigma=3$ one has $\sigma_{\rm NG}/\sigma_{\rm fs}-1 \approx 0.007$.

We expand the large-$t$ behavior (\ref{gptlarge}) of $G(t,\theta)$ as
\begin{equation}
G_P(t,\theta) \approx B_0 e^{- E_0 t} \left[ 1 + \theta^2 h(t) + ... \right]
\label{explt}
\end{equation}
where we set 
\begin{eqnarray}
&& B(\theta) = B_0 + \theta^2 B_2 + ... ,\\
&& E(\theta) = E_{0} + \theta^2 E_{2} + ...,
\end{eqnarray}
and
\begin{equation}
h(t) = \frac{B_2}{B_0} - E_{2} t
\end{equation}
Comparing the Eq.~(\ref{explt}) with the Eq.~(\ref{g2}), 
we find that
\begin{equation}
h(t) = {G^{(2)}_2(t)\over 2 G^{(0)}(t) }
\end{equation}
Thus $E_2$ can be estimated from the difference
\begin{equation}
\Delta h(t) = h(t)-h(t+a), 
\label{dft}
\end{equation}
indeed
\begin{equation}
{\displaystyle \lim_{t\to\infty}} \; \Delta h(t) = E_2\, a
\label{deltaht}
\end{equation}
Corrections are exponentially suppressed 
as $\exp[-(E^*_0-E_0) t]$ where $E^*_0$ is the mass of the first excited
state at $\theta=0$.  Assuming the free-string spectrum
(\ref{wnfree}), $E^*_0 - E_0= 4\pi/L$.  Notice that, although $E^*_0 -
E_0\rightarrow 0$ for $L\rightarrow \infty$, this difference is not
small in our calculations.  Indeed, since we choose the lattice size
$L$ so that $l_\sigma \equiv \sqrt{\sigma} L \approx 3$, 
$(E_0^*-E_0)/E_0 \approx 4\pi/l_\sigma^2 \approx 1.4$.

Finally, the coefficient $s_2$ of the $O(\theta^2)$ term in the
expansion (\ref{sigmaex}) is obtained by
\begin{equation}
s_2= {E_{2}\over \sigma L}
\label{s2ext}
\end{equation}
$s_2$ is a dimensionless scaling quantity. It is
expected to approach a constant in the continuum limit, with $O(a^2)$
scaling corrections.

An analogous procedure can be used to determine the leading
$O(\theta^2)$ term in the $\theta$ expansion of the lowest glueball
mass $M(\theta)$ around $\theta=0$, cf. Eq.~(\ref{gmex}).  In this case we
employ wall-wall correlators of plaquette-like operators with up to 6
links, all in spatial directions, in order to determine the $0^{++}$
glueball mass.  Correspondingly we define 
\begin{equation}
\Delta k(t) \equiv k(t)-k(t+a)
\label{deltas}
\end{equation}
where $k(t)$ is the function analogous to $h(t)$ defined
from the glueball wall-wall correlators. 
Then, the $O(\theta^2)$ coefficient $g_2$ in the expansion (\ref{gmex})
can be obtained by 
\begin{equation}
g_2 = {1\over a\, M} \,\lim_{t\to\infty} \; \Delta k(t) 
\label{g2ext}
\end{equation}
where $M$ is the $0^{++}$ glueball mass.

Finally, we mention that in order to improve the efficiency of the
measurements we used smearing and blocking procedures (see,
e.g., Refs.~\cite{smearing,LT-01}) to construct new operators with a better
overlap with the lightest propagating state.  Our implementation of
smearing and blocking was already described in Ref.~\cite{DPRV-02}. We
only mention that we constructed new super-links using three smearing,
and a few (2-4) blocking steps, according to the value of $L$. These
super-links were used to compute improved Polyakov lines or plaquette
operators.

\section{Results}
\label{res}

\TABLE[ht]{
\caption{ 
Some information on our Monte Carlo simulations for $N=3,4,6$.
The estimates of the string tension $\sigma$ are obtained
using Eq. (\protect\ref{lute}).
}
\label{tableruns}
\footnotesize
\begin{tabular}{cccclll}
\hline\hline
\multicolumn{1}{c}{$N$}&
\multicolumn{1}{c}{$\beta$}&
\multicolumn{1}{c}{lattice}&
\multicolumn{1}{c}{stat}&
\multicolumn{1}{c}{$a^2\,\sigma$}&
\multicolumn{1}{c}{$a\,M_{0^{++}}$}&
\multicolumn{1}{c}{$M_{0^{++}}/\sqrt{\sigma}$}\\
\hline \hline
3 & 5.9 & $12^3\times 18$ & 25M/20 & 0.0664(6)  & 0.80(1) & 3.09(4) \\

3 & 6.0 & $16^3\times 36$  & 25M/40 & 0.0470(3) & 0.70(1) & 3.23(4) \\

4 & 10.85 & $12^3\times 18$ & 16M/50 & 0.0646(6)& 0.76(1) & 2.99(5) \\

6 & 24.5  & $8^3\times 12$  & 9M/50 & 0.114(2)  & 0.83(1) & 2.46(4) \\

\hline\hline
\end{tabular}
}

In this section we present the results of our exploratory study of the
$\theta$ dependence of the spectrum using the method outlined in the preceding
section.  Table~\ref{tableruns} contains some information on our MC runs for
$N=3,4,6$ on lattices $L^3\times T$.  Since the coefficients of the $\theta$
expansions (\ref{sigmaex}) and (\ref{gmex}) are computed from connected
correlation functions, such as (\ref{g2}), and turn out to be quite small,
high statistics is required to distinguish their estimates from zero: Our runs
range from 9 to 25 million sweeps, with measures taken every 20-50 sweeps.
This requirement represents a serious limitation to the possibility of
performing runs for large lattices and in the continuum limit, especially for
large values of $N$, due also to the severe critical slowing down discussed in
Sec.~\ref{mc}. For all values of $\beta$ considered in this work, the
autocorrelation time satisfies $\tau_Q \lesssim 100$ sweeps~\cite{DPV-02}.
Furthermore, $\beta$ values were chosen to lie in the weak-coupling
region, i.e., beyond the first order phase transition in the case $N=6$, and
beyond the crossover region characterized by a peak of the specific heat for
$N=3,4$; see Ref.~\cite{DPRV-02} for a more detailed discussion of this point.
Runs generally started from cold configurations, to avoid problems due to
metastable states at the transition (in the case $N=6$).  The lattice size $L$
was chosen so that $l_\sigma\equiv \sqrt{\sigma}L\gtrsim 3$, which should be
sufficiently large to obtain infinite-volume results (see, e.g.,
Refs.~\cite{LT-01,DPRV-02}), at least within our precision.  Due to the
above-mentioned limitations, and in particular for $N=4,6$ we could afford
only one value of $\beta$, so that no stringent checks of scaling could be
performed. For this reason our study should be still considered as a first
exploratory investigation.

\FIGURE[ht]{

$\!\!\!\!\!\!\!$\epsfig{file=n3b5p9.eps, width=10truecm} 

\epsfig{file=n3b6.eps, width=10truecm} 
\caption{
Plot of $\Delta h(t)$ and $\Delta k(t)$ for $N=3$ at $\beta=5.9$ (above)
and $\beta=6.0$ (below).
The data for $\Delta k(t)$ are slightly shifted along the $t$ axis.
}
\label{n3}
}

Figs.~\ref{n3} and \ref{n46} show the results for the discrete differences
$\Delta h(t)$ and $\Delta k(t)$, cf. Eqs.~(\ref{dft}) and (\ref{deltas}), for
$N=3$ at $\beta=5.9,6.0$ and for $N=4,6$ respectively.  As
expected, the signal degrades rapidly with increasing $t$.  Anyway, they
appear rather stable already for small values of $t$. As already discussed in
Sec.~\ref{thex}, the approach to a constant in the large-$t$ limit should be
exponential, as $\exp[-(E^*_0-E_0) t]$, where $E^*_0$ is the energy of the
first excited state at $\theta=0$.  
In the case $N=3$ the data at $\beta=6$
appear to approach the asymptotic behavior more rapidly than at $\beta=5.9$.
This should be due to the fact that a more effective blocking procedure can be
applied when $L=16$, rather than $L=12$, achieving a better overlap with the
lowest state.

We estimate the coefficients $s_2$ and
$g_2$ of the $O(\theta^2)$ terms in the expansions (\ref{sigmaex}) and
(\ref{gmex}) from the corresponding discrete differences $\Delta h(t)$ and
$\Delta k(t)$ [cf. Eqs.~(\ref{deltaht}), (\ref{s2ext}), and (\ref{g2ext})],
taking the data at $t/a=2$ in the $N=3,4$ runs, and at $t/a=1$ for $N=6$.  In
Table~\ref{tableres} we report the results.  The estimates of $s_2$ and $g_2$
are small in all cases, and decrease with increasing $N$.  For $N=3$ the
results at $\beta=5.9$ and $\beta=6.0$ are consistent, supporting the expected
scaling behavior.  As final estimate one may consider
\begin{equation}
s_2=-0.08(1), \quad g_2=-0.06(2) \qquad {\rm for}\;\; N=3
\end{equation}

\FIGURE[ht]{

$\!\!\!\!\!\!\!$\epsfig{file=n4b.eps, width=10truecm} 

\epsfig{file=n6b.eps, width=10truecm} 
\caption{
Plot of $\Delta h(t)$ and $\Delta k(t)$ for
$N=4$ at $\beta=10.85$ (above) and $N=6$ at $\beta=24.5$ (below).
The data for $\Delta k(t)$ are slightly shifted along the $t$ axis.
}
\label{n46}
}

One may also consider the $\theta$ dependence of the scaling ratio
\begin{equation}
{M(\theta)\over \sqrt{\sigma(\theta)}} =
{M\over \sqrt{\sigma}} ( 1 +  c_2 \theta^2 + ... ),
\label{ratioex}
\end{equation}
where $c_2=g_2-s_2/2$. Using the numbers reported in Table~\ref{tableres},
we see that the $O(\theta^2)$ terms tend to cancel in the ratio.
Indeed, we find $c_2=-0.02(2),\, -0.01(3),\, -0.01(2)$ respectively 
for $N=3,4,6$.

For $N=4,6$ there are other
independent $k$-strings associated with representations
of higher $n$-ality. Analogously to the fundamental string,
one may write
\begin{equation}
\sigma_k(\theta) = \sigma_k \left( 1 + s_{k,2} \,\theta^2 + ... \right),
\label{sigmaexk}
\end{equation}
where $\sigma_k$ is the $k$-string tension at $\theta=0$.  
The case $k=1$ corresponds to the fundamental string tension,
i.e. $\sigma_1\equiv \sigma$ and $s_{1,2}\equiv s_2$.
One may also consider the ratio $R_k = \sigma_k/\sigma$,
\begin{equation}
R_k(\theta) = R_k \left( 1 + r_{k,2} \,\theta^2 + ... \right),
\label{rexk}
\end{equation}
where $R_k$ is the ratio at $\theta=0$ (see e.g.
Refs.~\cite{LT-01,DPRV-02,DPV-03,LTW-04} for recent numerical studies of the
$k$-string spectrum), and $r_{k,2}=s_{k,2}-s_2$.  In the case $N=4$ there is
one independent $k$ string, $\sigma_2$, besides the fundamental one; in the
case $N=6$ there are two.  Our results for the $k>1$ strings are less stable.
We obtained sufficiently precise results only from the simulation for $N=4$.
In the channel of Polyakov lines corresponding to the $k=2$ string, we found
$a^2\sigma_2=0.091(2)$, $\Delta h_2(t/a=1)=-0.044(13)$, while at distance
$t/a=2$ the signal was already unreliable. This leads to the estimate
$s_{2,2}=-0.040(12)$. Note that $s_{2,2}\approx s_{1,2}$ ($s_{1,2}\equiv s_2$
is reported in Table~\ref{tableres}), suggesting an even smaller $O(\theta^2)$
term in the ratio $R_2$, i.e. $|r_{2,2}|\lesssim 0.02$.

\hfill
\TABLE[ht]{
\caption{
Results for the coefficients $s_2$ and $g_2$, as derived from the discrete
differences at distance $t/a=2$ for $N=3,4$, and at $t/a=1$ for $N=6$.  
}
\label{tableres}
\footnotesize
\begin{tabular}{ccccll}
\hline\hline
\multicolumn{1}{c}{$N$}&&
\multicolumn{1}{c}{$\beta$}&&
\multicolumn{1}{c}{$s_2$}&
\multicolumn{1}{c}{$g_2$}\\
\hline \hline
3 &\qquad\qquad& 5.9  &\qquad\qquad&  $-$0.077(8)\qquad\qquad   & $-$0.05(2) \\

3 && 6.0  &&  $-$0.077(15)  & $-$0.07(4) \\

4 && 10.85 &&  $-$0.057(10) &  $-$0.04(3) \\

6 && 24.5  &&  $-$0.025(5) & $\phantom{-}$0.006(15) \\

\hline\hline
\end{tabular}
}\hfill

In conclusion, the above results show that the $O(\theta^2)$ terms in the
expansion around $\theta=0$ of the spectrum of SU($N$) gauge theories are very
small, especially when dimensionless ratios are considered.  Moreover, they
appear to decrease with increasing $N$, and the coefficients do not show
evidence of convergence to a nonzero value.  This is suggestive of a scenario
in which the $\theta$ dependence of the spectrum disappears in the large-$N$
limit, at least for sufficiently small values of $\theta$ around $\theta=0$.
General large-$N$ scaling arguments applied to the Lagrangian
(\ref{lagrangian}) indicate $\bar{\theta}\equiv \theta/N$ as the relevant
Lagrangian parameter in the large-$N$ limit \cite{Witten-80}.  In the case of
the spectrum, this would imply that $O(\theta^2)$ coefficients should decrease
as $1/N^{2}$.  This is roughly verified by our results, taking also into
account that they may be subject to scaling corrections, especially those at
$N=4,6$. For example, in the case of the string tension,
$s_2\approx s_{2,2}/N^2$ with $s_{2,2}\approx -0.9$\,.
 Of course, further investigations are required to put this scenario
on a firmer ground.

The hypothesis of a simple $\theta$ dependence in the large-$N$ limit may be
extended to finite temperature, up to the first-order transition point. Recent
studies~\cite{DPV-04,LTW-05} have shown that in the large-$N$ limit the
topological properties remain substantially unchanged up to the first-order
transition point.

\section{$\theta$ dependence in the two-dimensional CP$^{N-1}$ model}
\label{cpn}

Issues concerning the $\theta$ dependence can also be discussed in
two-dimensional CP$^{N-1}$ models~\cite{DDL-79,Witten-79b}, which are an
interesting theoretical laboratory. Indeed they present several
features that hold in QCD: asymptotic freedom, gauge invariance,
existence of a confining potential between non gauge invariant states
(that is eventually screened by the dynamical constituents), and
non-trivial topological structure (instantons, $\theta$ vacua).
Moreover, unlike SU($N$) gauge theories, a systematic $1/N$ expansion
can be performed around the large-$N$ saddle-point
solution~\cite{DDL-79,Witten-79b,CR-92}.

Analogously to four-dimensional SU($N$) gauge theories, one may add a $\theta$
term to the Lagrangian, writing
\begin{equation}
{\cal L}_\theta  = {N\over 2g} \overline{D_\mu z}\, D_\mu z +
i \theta {1\over2\pi}\,\epsilon_{\mu\nu}\, \partial_\mu A_\nu,
\label{lagrangiancpn}
\end{equation}
where $z$ is a $N$-component complex scalar field subject to the constraint
$\bar{z}z=1$, $A_\mu=i\bar{z}\partial_\mu z$ is a composite gauge field, and
$D_\mu =\partial_\mu +iA_\mu$ is a covariant derivative.  The topological
charge density is $q(x)={1\over2\pi}\,\epsilon_{\mu\nu}\, \partial_\mu A_\nu$
. Then one may study the $\theta$ dependence of the ground state and other
observables.  In the following we discuss this issue within the $1/N$
expansion, performed keeping $g$ fixed.  Simple large-$N$ scaling arguments
applied to the Lagrangian (\ref{lagrangiancpn}) indicate that the relevant
$\theta$ parameter in the large-$N$ limit
should be $\bar{\theta}\equiv \theta/N$.

As mass scale we consider the zero-momentum mass~\footnote{
This quantity is more suitable for a $1/N$-expansion than the
mass scale determined from the large-distance exponential decay of $G_P(x)$,
due to its analytical properties in $1/N$~\cite{CR-92}.} 
$M$ defined from the
small-momentum behavior of the Fourier trasformed two-point correlation
function of the operator $P_{ij}(x) \equiv \bar{z}_i(x) z_j(x)$,
\begin{equation} 
G_P(x-y) = \langle {\rm Tr}\, P(x) P(y) \rangle, \label{GP}
\end{equation}
i.e. from the relation 
\begin{equation}
\widetilde{G}_P(p)^{-1} = Z^{-1} [ M^2 + p^2 + O(p^4)]
\end{equation}
where $Z$ is a renormalization constant.

Analogously to SU($N$) gauge theories, the ground state energy
$F(\theta)$ depends on $\theta$. One may define a scaling ground state
energy $f(\theta)$ and expand it around $\theta=0$,
\begin{equation}
f(\theta) \equiv M^{-2} [F(\theta)-F(0)] 
= {1\over 2} C \theta^2 \left( 1 + \sum_{n=1} b_{2n} \theta^{2n} \right) 
\label{fthetacpn} 
\end{equation}
where $F(\theta)$ is defined as in Eq.~(\ref{GSE}), $M$ is the mass
scale at $\theta=0$, and $C,b_j$ are constants.  $C$ is the scaling
ratio $\chi/M^2$ at $\theta=0$, where $\chi$ is the topological
susceptibility, i.e. the two-point correlation function of the
topological charge density at zero momentum.  The correlation function
of the topological charge density, and in particular the topological
susceptibility, has been computed within the $1/N$ expansion
\cite{luescher-78,CR-91,v-99}.
We have 
\begin{equation}
C = \chi/M^2 = {1\over 2\pi N} + O(1/N^2)
\end{equation} 
The coefficients $b_{2n}$ are obtained from appropriate $2n$-point
correlation functions of the topological charge density operators at
$\theta=0$.  For example
\begin{eqnarray}
&&b_2 = - {\chi_4\over 12 \chi},\\
&&\chi_4 = {1\over V} \left[ 
\langle Q^4 \rangle_{\theta=0} - 3 \left( 
\langle Q^2 \rangle_{\theta=0} \right)^2 \right], \label{chi4}
\end{eqnarray}
where $Q=\int d^2 x \, q(x)$ is the topological charge.

We refer to Ref.~\cite{CR-92} for a discussion of the
$1/N$ expansion within CP$^{N-1}$ models, its set up, 
and the list of the corresponding Feynman rules.
In Fig.~\ref{b2dia} we show  the $1/N$-expansion
Feynman diagrams contributing to $\chi_4$ at leading order. 
The analysis of the Feynman diagrams of
the connected correlations necessary to compute $b_{2n}$
shows that they are suppressed in the
large-$N$ limit, as
\begin{equation}
b_{2n} = O(1/N^{2n}).
\end{equation}
This implies that the ground-state energy can be rewritten as
\begin{eqnarray} 
&&f(\theta) = N \bar{f}(\bar{\theta}\equiv \theta/N), 
\label{fthetabarcpn}\\ 
&& \bar{f}(\bar{\theta}) = 
{1\over 2} \bar{C} \bar{\theta}^2 
( 1 + \sum_{n=1} \bar{b}_{2n} \bar{\theta}^{2n} ), 
\nonumber
\end{eqnarray}
where $\bar{C}\equiv N C$ and $\bar{b}_{2n}=N^{2n}b_{2n}$ are $O(1)$ in
the large-$N$ limit.  Note the analogy with the expected $\theta$ dependence
of the ground-state energy in SU($N$) gauge theories, cf.
Eq.~(\ref{fthetabar}).
The calculation of the coefficients of the leading large-$N$ terms is
rather cumbersome.  
Here, we only report the results obtained for
the leading terms of $\bar{b}_2$ and $\bar{b}_4$ 
\begin{equation}
\bar{b}_2= - {27\over 5}, \qquad \bar{b}_4= -{1830\over 7}.
\end{equation}

Within the $1/N$ expansion one may also study the dependence of the mass
$M$ on the parameter $\theta$. We write
\begin{equation}
M(\theta) = M\left( 1 + m_2 \theta^2 + ... \right)
\label{gmexcpn}
\end{equation} 
The coefficient $m_2$ can be extracted from the connected correlations
\begin{equation}
\langle {\rm Tr}\, P(x) P(0) Q^2 \rangle_{\theta=0} -
\langle {\rm Tr}\, P(x) P(0) \rangle_{\theta=0}
\langle Q^2 \rangle_{\theta=0}
\end{equation}
The analysis of its diagrams giving the corresponding
$1/N$ expansion indicates that $m_2$ is suppressed as
\begin{equation}
m_2 = O(1/N^2)
\end{equation}
This confirms the general large-$N$ scaling arguments indicating
$\bar{\theta}\equiv \theta/N$ as the relevant parameter in the
large-$N$ limit, as in the scenario put forward for the
four-dimensional SU($N$) gauge theories.

\FIGURE[ht]{
\epsfig{file=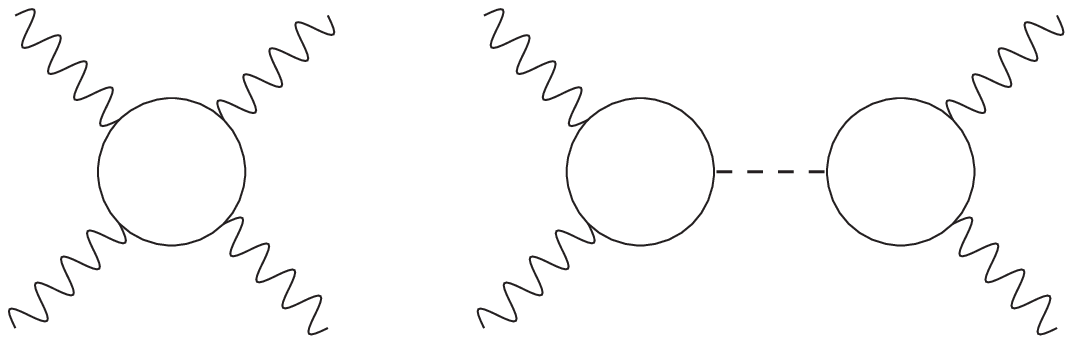, width=12truecm} 
\caption{
Diagrams contributing to the four-point connected function
$\chi_4$, cf. Eq.~(\ref{chi4}), within the $1/N$ expansion
(the corresponding  Feynman rules can be found in Ref.~\cite{CR-92}).
}
\label{b2dia}
}

\acknowledgments{We thank Michele Caselle, Martin Hasenbusch and
  Herbert Neuberger for helpful
  discussions. This work is supported in part by the 
Research Promotion Foundation of Cyprus (Proposal Nr: $\rm ENTA\Xi$/0504/11).}

\end{document}